\documentclass[12pt, journal, onecolumn]{IEEEtran}

\newif\iffull
\fulltrue
\usepackage{graphicx}
\usepackage{subfigure}
\usepackage{algorithmic, algorithm}
\usepackage[english]{babel}
\usepackage{amssymb}
\usepackage{amsmath}
\usepackage{cite}
\usepackage{graphicx}
\usepackage{amsthm}
\usepackage{thmtools, thm-restate}
\usepackage[hidelinks]{hyperref}
\usepackage[letterpaper,margin=0.8in]{geometry}
\usepackage[T1]{fontenc} 
\usepackage[english]{babel} 
\usepackage{amsmath,amsfonts,amsthm,amssymb} 

\usepackage{lipsum} 
\usepackage{amsmath, amssymb}
\usepackage[dvipsnames]{xcolor} 
\usepackage{microtype}
\usepackage{color}
\usepackage{cite}
\usepackage{pifont}
\usepackage{enumitem}
\usepackage{graphicx}
\usepackage[dvipsnames]{xcolor}
\usepackage{mathtools}
\usepackage{cleveref}
\usepackage{verbatim}
\usepackage[utf8]{inputenc}
\usepackage[english]{babel}

\theoremstyle{definition}
\newtheorem{theorem}{Theorem}
\newtheorem{lemma}{Lemma}
\newtheorem{claim}{Claim}
\newtheorem{definition}{Definition}
\newtheorem{corollary}{Corollary}

\newtheorem{construction}{Construction}
\newtheorem{example}{Example}
\newtheorem{observation}{Observation}

\newcommand{\bfa}{{\boldsymbol a}}

\newcommand{\bfc}{{\boldsymbol c}}

\newcommand{\bfs}{{\boldsymbol s}}

\newcommand{\bfu}{{\boldsymbol u}}

\newcommand{\bfx}{{\boldsymbol x}}
\newcommand{\bfy}{{\boldsymbol y}}
\newcommand{\bfz}{{\boldsymbol z}}

\newcommand{\indlen}{$\ell$ }
\newcommand{\nodollarindlen}{\ell}
\newcommand{\NameCodes}{DNA-correcting code}
\newcommand{\MatrixCorrecting}{$(\nodollarindlen,M,d)$-matrix}
\newcommand{\lMatCorrecting}{$(\nodollarindlen',M,d)$-matrix}
\newcommand{\indexCode}{$(\nodollarindlen,M,d)$ index-correcting code}
\newcommand{\indexLogMCode}{$(\log(M),M,d)$ index-correcting code}

\newcommand{\bijection}{BI}
\newcommand{\BoldData}{\bfu}

\newcommand{\cC}{\mathcal{C}}
\newcommand{\cD}{\mathcal{D}}

\newcommand{\cG}{\mathcal{G}}

\newcommand{\cX}{\mathcal{X}}

\newcommand{\cZ}{\mathcal{Z}}

\newcommand{\db}[1]{{\footnotesize  [{\textcolor{purple}{#1}} \textcolor{purple!60!black}{--Daniella}]\normalsize}}

\newcommand{\ab}[1]{{\footnotesize  [{\textcolor{blue}{#1}} \textcolor{blue!60!black}{--avital}]\normalsize}}

\title{DNA-Correcting Codes: End-to-end Correction in DNA Storage Systems}

\author{%
   \IEEEauthorblockN{\textbf{Avital~Boruchovsky}, 
                     \textbf{Daniella~Bar-Lev} \textbf{Eitan~Yaakobi}}
   \IEEEauthorblockA{\\Department of Computer Science, 
                  Technion---Israel Institute of Technology, 
                     Haifa 3200003, Israel}
                      Email: \{avital.bor ,daniellalev, yaakobi\}@cs.technion.ac.il
   
   \thanks{Parts of this work were presented at the IEEE International Symposium on Information Theory (ISIT), Taipei, Taiwan, 2023~\cite{BBY23}.}
  \thanks{A. Boruchovsky, D. Bar-Lev, E. Yaakobi are with the Henry and Marilyn Taub Faculty of Computer Science, Technion - Israel Institute of Technology, Haifa 3200003, Israel 
  (e-mail: \texttt{avital.bor@campus.technion.ac.il ,daniellalev@cs.technion.ac.il, yaakobi@cs.technion.ac.il}).}
   \thanks{%
 The research was Funded by the European Union (ERC, DNAStorage, 865630). Views and opinions expressed are however those of the author(s) only and do not necessarily reflect those of the European Union or the European Research Council Executive Agency. Neither the European Union nor the granting authority can be held responsible for them.
}%
 }

\IEEEoverridecommandlockouts
\IEEEaftertitletext{\vspace{-5pt}}

\begin{document}
\maketitle
\begin{abstract}

This paper introduces a new solution to DNA storage that integrates all three steps of retrieval, namely clustering, reconstruction, and error correction. \emph{DNA-correcting codes} are presented as a unique solution to the problem of ensuring that the output of the storage system is unique for any valid set of input strands. To this end, we introduce a novel distance metric to capture the unique behavior of the DNA storage system and provide necessary and sufficient conditions for DNA-correcting codes. The paper also includes several bounds and constructions of DNA-correcting codes.
\end{abstract}


\section{Introduction}
\label{sec:introduction}

The first two experiments that showed the potential of using synthetic DNA as a means for a large-scale information storage system were done in\cite{CYK12} and\cite{GBCDLSB13}. Since then, together with the developments in synthesis and sequencing technologies, more research groups showed the potential of in vitro DNA storage; see e.g. \cite{AVAAY19,BOSEY21, BGHCTIP16,BLCGSS16,EZ17,O17,YGM17,YYMZM15}.

A typical DNA storage system consists of three components: (1) synthesization of the strands that contain the encoded data. In current technologists, each strand has millions of copies, and the length of these strands is usually limited to 250-300 nucleotides; (2) a storage container that stores the synthetic DNA strands; (3) a DNA sequencer that reads the strands, where the output sequences from the sequencing machine are called \emph{reads}. This novel technology has several properties that are fundamentally different from its digital counterparts, while the most prominent one is that the erroneous copies are stored in an unordered manner in the storage container (see e.g. \cite{HMG19}). The most common solution to overcome this challenge is to use indices that are stored as part of the strand. Each strand is prefixed with some nucleotides that indicate the strand’s location, with respect to all other strands. These indices are usually protected using an \emph{error-correcting code} (ECC) \cite{BOSEY21}, \cite{BGHCTIP16}, \cite{GHPPS15}, \cite{O17},\cite{YGM17}. The retrieval of the input information is usually done by the following three steps. The first step is to partition all the reads into \emph{clusters} such that the reads at each cluster are all noisy copies of the same information strand. The second step applies a \emph{reconstruction algorithm} on every cluster to retrieve an approximation of the original input strands. In the last step, an ECC is used in order to correct the remaining errors and retrieve the user's information.

While previous works tackled each of these steps independently (see e.g. \cite{AVAAY19},\cite{BOSEY21}, \cite{BGHCTIP16},\cite{O17}, \cite{SGPY21},  \cite{YGM17}), this work aims to tackle all of them \emph{together}. This is accomplished by limiting the stored messages in the DNA storage system, such that for any two input messages, the sets of all the possible outputs  will be mutually disjoint. We call this family of codes \emph{\NameCodes s}. Our point of departure is the recent work~\cite{SYLWZ20} of \emph{clustering-correcting codes} that proposed codes for successful clustering. However, their results have been established under the assumption that the correct reads in every  cluster satisfy some dominance property. Furthermore, the codes in~\cite{SYLWZ20} do not aim to recover the input data, but only to achieve a successful clustering. On the contrary, our suggested codes also guarantee the recovery of the input data, while eliminating the dominance assumption.   
Similar to~\cite{SYLWZ20}, it is assumed that every information strand consists of an \emph{index-field} and a \emph{data-field}.   

The rest of the paper is organized as follows. Section \ref{sec:defs} presents the definitions and the problem statement. In Section \ref{sec:error_free}, we consider the case where the data-field is error-free. In addition, we present the DNA-distance metric, which is used in order to find necessary and sufficient conditions for \NameCodes s. In section \ref{Sec:constructionSection}, we study codes over the index-field. Using these codes we present constructions for \NameCodes s and bounds on the size of  \NameCodes s. Lastly, in Section~\ref{Sec:ErrornesData} we generalize the results for the case of an erroneous data-field. 

\section{Definitions, Problem Statement, and Related Works}\label{sec:defs}

\subsection{Definitions}
The following notations will be used in this paper. For a positive integer $n$, the set $\{0,1,\dots,n-1\}$ is denoted by $[n]$ and $\{0,1\}^n$ is the set of all length-$n$ binary vectors. For two vectors $\bfx,\bfy,$ of the same length, the Hamming distance between them is the number of coordinates in which they differ and is denoted by $d_H(\bfx,\bfy)$. 
For two sets of vectors of the same size $Z,Y$, let $\bijection(Z,Y)$ be the space of all bijective functions (matchings) from $Z$ to $Y$. For a matching $\pi\in \bijection(Z,Y)$, let $w_H(\pi)$ denote the maximal Hamming distance between any two matched vectors, i.e., $w_H(\pi)=\max\limits_{\bfz \in Z}\{d_H(\bfz,\pi(\bfz))\}$. 
We assume a binary alphabet in the paper as the generalization to higher alphabets will be immediate and all logs are taken according to base 2. 

Assume that a set of $M$ unordered length-$L$ strands are stored in a DNA-based storage system. We will assume that $M=2^{\beta L}$ for some $0<\beta<1$, and for simplicity, it is assumed that $\beta L$ is an integer. Every stored length-$L$ strand $\bfs$ is of the form $\bfs=(ind,\bfu)$, where $ind$ is the length-$\bf{\nodollarindlen}$ index-field of the strand (which represents the relative position of this strand in relation to all other strands) and $\bfu$ is the length-$(L-\nodollarindlen)$ data-field of the strand. Different strands are required to have a different index-field, as otherwise, it will not be possible to determine the order of the strands. The length of the index-field of all the strands is the same and since all indices are different it holds that \indlen $\geq \log(M)=\beta L$.

For  $M,L$, and $\ell$, the space of all possible messages that can be stored in the DNA storage system is:
\begin{align}\label{XMLl}
   \cX_{M,L,\nodollarindlen}=\big\{\{&(ind_1,\bfu_1),(ind_2,\bfu_2),\dots,(ind_{M},\bfu_{M})\}| \\
& \forall i: ind_i\in \{{0,1}\}^{\nodollarindlen}, \bfu_i\in \{0,1\}^{L-\nodollarindlen}, \forall i\neq j:
ind_i\neq ind_j  \big\}\nonumber .
\end{align}
Note that  $|\cX_{M,L,\nodollarindlen}|= {2^\nodollarindlen \choose M}{2^{(L-\nodollarindlen)M}}$ since there are $2^\nodollarindlen \choose M$ options to choose the different set of index-fields and then $2^{(L-\nodollarindlen)M}$ more options to choose the data-field for every index. Under this setup, a code {$\cC$} will be a subset of {$\cX_{M,L,\nodollarindlen}$}.

\subsection{Problem Statement}
When a set $Z=\{(ind_1,\bfu_1),\dots,(ind_{M},\bfu_{M})\}$ is synthesized, each of its strands has a large number of noisy copies, and during the sequencing process a subset of these copies is read, while the number  of copies mostly depends on the synthesis and sequencing technologies. Throughout this paper, it is assumed that the number of copies for each strand is exactly $K$, and so, the sequencer's output is a set of $MK$ reads, where every output read is a noisy copy of one of the input strands. It is also assumed that the noise is of substitution type and in Section \ref{sec:conclusions} we explain how most of the results hold for edit errors as well when changing the Hamming distance to the edit distance. 
Let {$\tau$} denote the maximal relative fraction from the $K$ reads of each input strand that can be erroneous. Furthermore, let $e_i,e_d$ be the largest number of errors that can occur at the index-field, data-field of each strand, respectively. Formally, the DNA storage channel is modeled as follows.

\begin{definition}
A DNA-based storage system is called a \emph{$(\tau,e_i,e_d)_K$-DNA storage system} if it satisfies the following properties: 
(1) Every input strand $(ind,\bfu)$ has exactly $K$ output reads, (2) at most $\lfloor \tau K\rfloor $ of these reads are erroneous, and (3) if $(ind',\bfu')$ is a noisy read of $(ind,\bfu)$ then  $d_H(ind,ind')\leq e_i$ and $d_H(\bfu,\bfu')\leq e_d$.
\end{definition}

For a set $Z\in \cX_{M,L,\nodollarindlen}$, let $B^K_{(\tau,e_i,e_d)}(Z)$ be the set of all possible $MK$ reads one can get from {$Z$} after it passes through a {$(\tau,e_i,e_d)_K$}-DNA storage system (i.e., every element in $B^K_{(\tau,e_i,e_d)}(Z)$ is a multiset of $MK$ reads).
Under this setup, a code {$\cC\subseteq \cX_{M,L,\nodollarindlen}$} is called a\ \emph{{$(\tau,e_i,e_d)_K$}-\NameCodes} if for every two codewords {$Z,Z'\in \cC $} such that {$Z\neq Z' $}, it holds that
{$B^K_{(\tau,e_i,e_d)}(Z) \cap B^K_{(\tau,e_i,e_d)}(Z')=\emptyset  $}, i.e., the sets of possible outputs for all codewords are mutually disjoint when the parameters are {$\tau,e_i,e_d$} and $K$. 
The redundancy of such a code is defined by {$r(\cC)=\log_2(|\cX_{M,L,\nodollarindlen}|)-\log_2(|\cC|)$}. 

Let $A_{M,L,\nodollarindlen}(\tau,e_i,e_d,K)$ denote the largest size of a {$(\tau,e_i,e_d)_K$}-\NameCodes\ given the parameters $M,L,\nodollarindlen,\tau,e_i,e_d,$ and $K$. The goal of this work is to find necessary and sufficient conditions for a code to be \NameCodes\ and study the value of $A_{M,L,\nodollarindlen}(\tau,e_i,e_d,K)$, for different parameters. 

\subsection{Related Work}\label{Related Work}
Previous studies on information retrieval in DNA storage systems have typically tackled the problem by addressing the three steps (i.e., clustering, reconstruction, and error correction) individually, utilizing a combination of ECC and algorithmic methods.
In most works,  the clustering step was performed by protecting the indices with an ECC and then using the decoder of this code to correct the indices and cluster the reads \cite{BGHCTIP16}, \cite{GHPPS15}, \cite{O17}, \cite{YGM17}. Other works used algorithmic methods which are usually time-consuming 
or not accurate enough in clustering \cite{QYW22},\cite{RMRAJYCS17}. The reconstruction task is commonly studied independently, and it is usually assumed that the clustering step was successful \cite{BMALA}, \cite{SYSY20}, \cite{SGPY21}. Additionally, in most previous works, an ECC is applied on the data and is used for correcting errors on the reconstructed strands, see e.g. \cite{BOSEY21}, \cite{BGHCTIP16}, \cite{O17},\cite{YGM17}. 
Another approach appears in~\cite{SYLWZ20}, where the authors studied the clustering problem from a coding theory perspective, however, their work only tackles the first step in the retrieval process of the data, i.e., the clustering step. 

The key difference of this work from previous studies is that in this work we present a novel approach for error-correcting codes in DNA storage systems which encapsulate all the information retrieval steps together into a single code.  
A follow-up work to the first version of this paper~\cite{BBY23} appears in \cite{HWU23}. In~\cite{HWU23}, Wu generalized some of the results that are presented in Section~\ref{sec:error_free} to the case where errors can also affect the data-field. For more details see Section~\ref{Sec:ErrornesData}.

\section{Error Free Data-Field }\label{sec:error_free}
We start by studying the case where the data part is free of errors, i.e., $e_d=0$.
For a set $Z=\{(ind_1,\bfu_1),\dots,(ind_{M},\bfu_{M})\}\in \cX_{M,L,\nodollarindlen}$, let $S(Z)$ denote \emph{the data-field set} of $Z$ which is defined by  $S(Z) = \{\bfu_1,\dots,\bfu_{M}\}$ and $MS(Z)$ denotes \emph{the data-field multiset} of $Z$, $MS(Z) =\{\!\{\bfu_1,\dots,\bfu_{M}\}\!\}$. We use the notation of $MS(\cX_{M,L,\nodollarindlen})$ to denote the set of all possible data-field multisets of the elements in $\cX_{M,L,\nodollarindlen}$.

For a code $\cC\subseteq \cX_{M,L,\nodollarindlen}$ and a data-field multiset $U\in MS(\cX_{M,L,\nodollarindlen})$, let ${\cC_U\subseteq\cC}$ be the set of all codewords  $Z\in\cC$ for which $MS(Z) =U$.
The next claim presents a necessary and sufficient condition for DNA-correcting codes for $e_d=0$.

\begin{claim}\label{cl:basic}
A code $\cC\subseteq \cX_{M,L,\nodollarindlen}$ is a {$(\tau,e_i,e_d = 0)_K$}-DNA-correcting code if and only if for every data-field multiset $U\in MS(\cX_{M,L,\nodollarindlen})$, it holds that $\cC_U$ is a {$(\tau,e_i,0)_K$}-\NameCodes.  
\end{claim}

\begin{IEEEproof}
If $\cC\subseteq \cX_{M,L,\nodollarindlen}$ is a {$(\tau,e_i,0)_K$}-\NameCodes, then every subset of it is a {$(\tau,e_i,0)_K$}-\NameCodes\ as well. On the other hand, if $Z_1,Z_2 \in \cC$ such that $MS(Z_1)\neq MS(Z_2)$ then $B^K_{(\tau,e_i,0)}(Z_1) \cap B^K_{(\tau,e_i,0)}(Z_2)=\emptyset$ since the data-field is free of errors, and for $Z_1,Z_2 \in\cC$ such that $MS(Z_1)= MS(Z_2)$ we have that $B^K_{(\tau,e_i,0)}(Z_1) \cap B^K_{(\tau,e_i,0)}(Z_2)=\emptyset$, since $\cC_{MS(Z_1)}$ is a {$(\tau,e_i,e_d)_K$}-\NameCodes.
\end{IEEEproof}

For a data-field multiset $U\in MS(\cX_{M,L,\nodollarindlen})$, let $A_{M,L,\nodollarindlen}(\tau,e_i,e_d,K)_U $ denote the 
largest size of a {$(\tau,e_i,e_d)_K$}-\NameCodes\ in which all its codewords have a data-field multiset $U$. The next corollary follows immediately from  Claim \ref{cl:basic}.

\begin{corollary}\label{corollary-codeForDifferentMultiSets}
It holds that
$${A_{M,L,\nodollarindlen}(\tau,e_i,0,K) =\sum\limits_{U\in MS(\cX_{M,L,\nodollarindlen})} A_{M,L,\nodollarindlen}(\tau,e_i,0,K)_U}.$$
\end{corollary}

The last corollary implies that for $e_d=0$, in order to find the largest \NameCodes\ it is sufficient to find the largest \NameCodes\ for every data-field multiset $U$. To this end, we define the DNA-distance, a metric on ${\cX}_{M,L,\nodollarindlen}$, which will be useful for determining what conditions a {$(\tau,e_i,0)_K$}-\NameCodes\ $\cC_U$ must hold.

\subsection{The DNA-Distance}
For $Z = \{(ind_1,\bfu_1),\dots,(ind_{M},\bfu_{M})\}\in {\cX}_{M,L,\nodollarindlen}$ and $\bfu\in S(Z)$, let $I(\bfu,Z)$ be the set of all indices of $\bfu$ in $Z$, that is, $I(\bfu,Z) = \{ind_i | \ \bfu_i = \bfu\}$. For $Z_1,Z_2 \in \cX_{M,L,\nodollarindlen}$, their \emph{DNA-distance} is defined as
\begin{equation*}
\cD(Z_1,Z_2)  = 
\begin{cases}
\infty,   & \text{\hspace{-0ex}if }  MS(Z_1) \neq  MS(Z_2), \\
\hspace{-0.5ex}\max\limits_{\scalebox{0.7}{$\bfu\hspace{-0.5ex}\in\hspace{-0.5ex} S(Z_1)$}} \min\limits_{\scalebox{0.7}{$\hspace{0.5ex}\pi\in\hspace{-0.5ex} BI(I{(\bfu,Z_1)},I{(\bfu,Z_2)})$}}\{w_H(\pi)\},  &\text{otherwise.}
\end{cases} 
\end{equation*}
That is, if the data-field multisets are different, then the distance is infinity. Otherwise, for each data-field $\bfu$ we look at all the possible matchings between the two sets of indices of $\bfu$ in $Z_1$ and $Z_2$, and choose the matching with minimal Hamming weight. Then we take the maximal data-field $\bfu$ and the distance is the Hamming weight of the minimal matching for this data-field. The motivation for using the DNA-distance is Claim~\ref{cl:basic} and the observation that if the codewords $Z_1$ and $Z_2$ have different data-field multisets then they cannot share the same output. However, if their data-field multisets are the same, then we consider the Hamming distance between the index-fields of the same data-field. Given a set {$Z\in {\cX}_{M,L,\nodollarindlen}$}, we define the radius-$r$ ball\footnote{We use the terminology of a ball since $\cD$ is a metric, as shown in Lemma~\ref{lemma:distance-matric}.} of {$Z$} by {$B_r(Z)=\{Y\in {\cX}_{M,L,\nodollarindlen} \ | \ \cD(Z,Y)\leq r\}$}.

\begin{example}
 Consider the following two words in ${\cX}_{4,5,2}$ 
\begin{align*}
&Z_1=\{(00,111),(01,000),(10,111),(11,001)\},\\
&Z_2=\{(00,111),(01,111),(10,001),(11,000)\}.
\end{align*}
 Both words have the same data-field multiset and thus the DNA-distance between them is not infinity. In Fig.~\ref{fig:dist_ex}, for every $\BoldData\in S(Z_1)$, we show all possible matchings between $I(\BoldData,Z_1)$ and $I(\BoldData,Z_2)$.
\begin{figure}
     \centering   \includegraphics[width=0.5\textwidth]{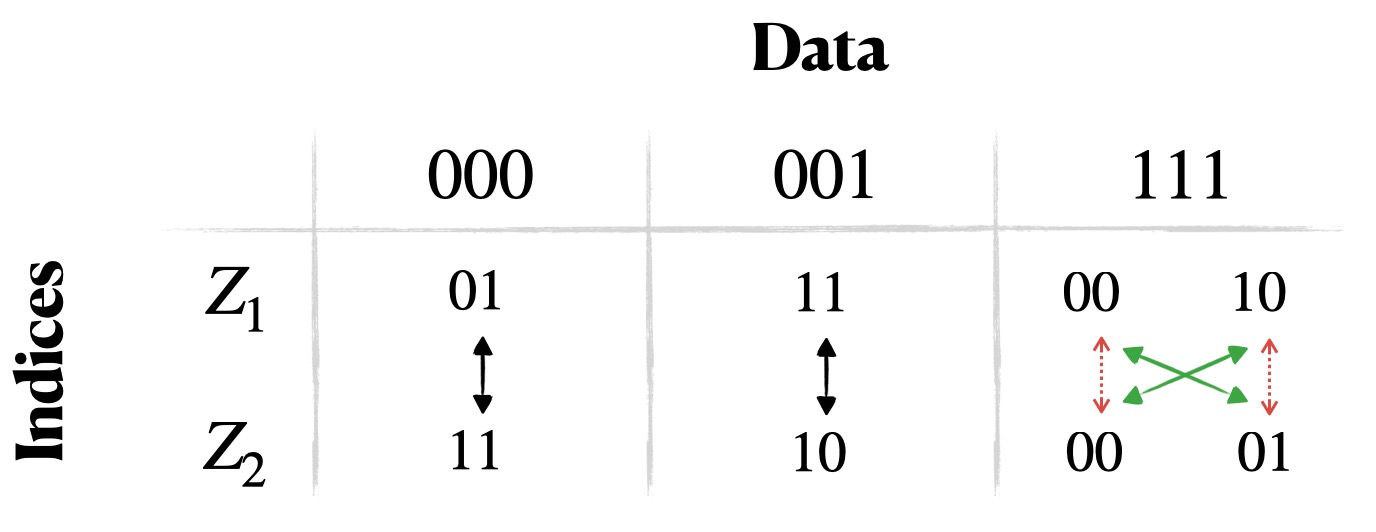}
     \caption{All possible  matchings between $I(\BoldData,Z_1)$ and $I(\BoldData,Z_2)$ for every date field $\BoldData\in S(Z_1).$}
     \label{fig:dist_ex}
     \end{figure}
The data-fields $000$ and $001$ have only one index in both $Z_1$ and $Z_2$. Thus, there is only one matching in both cases and the weight of each matching is one. On the other hand, the data-field $111$ has two indices in $Z_1$ and $Z_2$, and thus there are two optional matchings between the corresponding indices in $Z_1$ to the corresponding indices in $Z_2$, the dashed red one and the solid green as can be seen in Figure~\ref{lemma:distance-matric}. The weight of the red matching is 2 since  $w_H(red)=\max\{d_H(00,00),d_H(10,01)\}=2$ and the weight of the green matching is 1 since $w_H(green)=\max\{d_H(00,01),d_H(10,00)\}=1$. Hence, $\cD(Z_1,Z_2)=1$. 
\end{example}

As will be seen later, the DNA-distance $\cD$ will be essential in order to determine if a code is a \NameCodes. First, we show that $\cD$ is a metric but not a graphic metric\footnote{A metric $\cD:X\times X\rightarrow \mathbb{N}$ is graphic if the graph $G=(V,E)$ with $V=X$ and edges connect between any two nodes of distance one, satisfies the following property: for $x_1,x_2\in X$ it holds that $\cD(x_1,x_2)=t$ if and only if the length of the shortest path between $x_1$ and $x_2$ in $G$ is $t$ as well.}. 

\begin{restatable}{lemma}{distancemetric}
\label{lemma:distance-matric}
The DNA-distance $\cD$ is a metric on {$ {\cX}_{M,L,\nodollarindlen}$}, but not a graphic one.
\end{restatable}
\begin{IEEEproof}
Let $Z_1,Z_2,Z_3\in {\cX}_{M,L,\nodollarindlen}$. In order to show that $\cD$ is indeed a metric, let us prove the three properties of a metric:
\begin{enumerate}
\item {$\cD(Z_1,Z_2)=0 \text{ if and only if }  Z_1=Z_2$.}

If {$ Z_1=Z_2$} then for every data-field $\BoldData$ we consider the trivial matching that sends every index to itself, hence {$\cD(Z_1,Z_2)=0 $}. For the other direction, if {$\cD(Z_1,Z_2)=$~$0$} then both words have the same data-field multiset, and in the minimal matching for every data-filed, every index is matched to itself (otherwise the distance would be greater than $0$), thus the indices for every data-field $\BoldData$ are the same in both sets and hence {$ Z_1=Z_2$}.

\item {$\cD(Z_1,Z_2)=\cD(Z_2,Z_1)$.}

Since there is no difference between matching the indices in $Z_1$ and the indices in $Z_2$ or vice versa, the definition of the distance is symmetric. 

\item {$\cD(Z_1,Z_3)\leq \cD(Z_1,Z_2)+\cD(Z_2,Z_3)$.}

If {$ Z_1 \text{ and } Z_3$} do not have the same data-field multiset then at least {$ Z_1\text{ and } Z_3$}  or {$ Z_2 \text{ and } Z_3$} have different data-field multiset, and the right-hand side will be equal to  {$\infty$} as well. Assume that $Z_1, Z_2, Z_3$ have the same data-field multiset $\{\!\{\bfu_1,\bfu_2,\dots,\bfu_M\}\!\}$. Let $\pi_{i,j}$ denote the matching between the indices of $Z_i$ and $Z_j$, where every index  of data-field $\bfu$ is matched according to the minimal Hamming weight matching in $BI(I{(\bfu,Z_i)},I{(\bfu,Z_j)})$.
We show that for every $\bfu$ the triangle inequality holds and our claim will follow. Let there be $\bfu\in S(Z_1) $ and let $\cD(Z_1,Z_2)_\bfu$ be the distance when focusing on the data-field $\bfu$. Then, it holds that  
\begin{align*}
\cD(Z_1,Z_2)_\BoldData+\cD(Z_2,Z_3)_\BoldData&=\max\limits_{i \in I(\BoldData,Z_1)}d_H (i,\pi_{1,2}(i))+\max\limits_{i\in I(\BoldData,Z_1)}d_H\big(\pi_{1,2}(i),\pi_{2,3}(\pi_{1,2}(i))\big) \\
&\geq\max\limits_{i\in I(\BoldData,Z_1)}\{d_H (i,\pi_{1,2}(i))+d_H\big(\pi_{1,2}(i),\pi_{2,3}(\pi_{1,2}(i))\big)\}\\
& \overset{\mathrm{(a)}}{\geq} \max\limits_{i\in I(\BoldData,Z_1)}\{d_H(i,\pi_{23}(\pi_{12}(i))\}\\&\overset{\mathrm{(b)}}\geq  \cD(Z_1,Z_3)_\BoldData.
\end{align*}

where (a) holds since the Hamming distance is a metric and (b) is true because $\pi_{23}\circ\pi_{12}\in \bijection(I(\BoldData,Z_1),I(\BoldData,Z_3))$.
\end{enumerate}

Next, we show by a counter-example that the metric is not graphic.
Let there be the following two sets in $\cX_{4,L,4}$
\begin{align*}
&Z_1=\{(0000,\BoldData_1),(1100,\BoldData_2),(1010,\BoldData_3),(1001,\BoldData_4)\},\\
&Z_2=\{(0000,\BoldData_2),(1100,\BoldData_1),(1010,\BoldData_4),(1001,\BoldData_3)\},
\end{align*}
where all the data-fields are different. On one hand, since for every data-field, the Hamming distance between the corresponding indices is 2 it holds that $\cD(Z_1,Z_2)=2$. On the other hand, if the metric was graphic, the shortest path between $Z_1$ and $Z_2$ in the corresponding graph would be 2 as well, which implies that there would be a word $Z \in \cX_{4,L,4}$ such that {$Z\in B_{1}(Z_1)\cap B_{1}(Z_2)$}. But for it to happen, $Z$ must have the same data-field multiset as $Z_1$ and $Z_2$, and the indices of {$\BoldData_1$} and $\BoldData_2$ must be {$\{1000,0100\}$}, and the indices of  {$\BoldData_3$} and $\BoldData_4$ must be {$\{1000,1011\}$}. We get three indices for four data parts, and hence $B_{1}(Z_1)\cap B_{1}(Z_2)=\emptyset$. 
\end{IEEEproof}

Even though the DNA-distance is not a graphic metric, it still satisfies several properties that hold trivially for such ones. In particular, using the metric $\cD$ it is possible to derive necessary and sufficient conditions for a code to be a \NameCodes, which are shown in the next subsection. 

\subsection{Necessary and Sufficient Conditions for DNA-Correcting Codes} 

For a code $\cC\subseteq {\cX}_{M,L,\nodollarindlen}$, its \emph{DNA-distance} is defined by $\cD(\cC) \triangleq \min\limits_{Z_1\neq Z_2\in \cC} \cD(Z_1,Z_2)$.
Next, we draw connections between DNA-correcting codes and their DNA-distance. These connections will depend upon the value of $\tau$. First, the case $\tau=1$ is considered. In the proof of Theorem \ref{theorem-tau=1}, we use Hall's marriage theorem, which is stated next.

\begin{theorem}[\emph{Hall, 1935}]
For a finite bipartite graph $G=(L\cup R,E)$, there is an $L$-perfect matching\footnote{In a bipartite graph $G=(L\cup R,E)$, an $L$ perfect matching is a subset $T$ of the edges $E$, such that every vertex in $L$ is adjacent to exactly one edge in $T$.} if and only if for every subset $Y\subseteq L$ it holds that $|Y|\leq |N_G(Y)|$, where $N_G(Y)$ is the set of all vertices  that are adjacent to at least one element of $Y$.   
\end{theorem}
\begin{theorem}\label{theorem-tau=1}
 A code {$\cC\subseteq {\cX}_{M,L,\nodollarindlen}$} is a { {$(1,e_i,0)_K$}-\NameCodes} if and only if $\cD(\cC)>2e_i$.

\end{theorem}
\begin{IEEEproof}
From Claim \ref{cl:basic} it is sufficient to show that the claim holds for every $\cC_U\subseteq \cC$. Let $U=\{\!\{\BoldData_1,\BoldData_2,\dots,\BoldData_M\}\!\}\in MS(\cX_{M,L,\nodollarindlen})$ and assume that {$\cC_U\subseteq {\cX}_{M,L,\nodollarindlen}$} is a {{$(1,e_i,0)$}-\NameCodes}. Assume to the contrary that there are two codewords {$ Z_1,Z_2\in \cC_U$} such that {$\cD(Z_1,Z_2)\leq2e_i$}. It will be shown that there exists {$W\in B^K_{(1,e_i,0)}(Z_1)\cap B^K_{(1,e_i,0)}(Z_2)$}. For every data-field {${\BoldData_i\in S(Z_1)}\hspace{0.04EX}$}, there exists $\pi_i\in\bijection\left((I(\BoldData_i,Z_1),I(\BoldData_i,Z_2))\right)$ such that $w_H(\pi_i)\leq2e_i$. Thus, for every index $i_j\in I(\BoldData_i,Z_1)$ there exists $r_{i_j}\in {\{0,1\}}^\nodollarindlen$ with $d_H(i_j,r_{i_j})\leq e_i$ and $d_H(\pi_i(i_j),r_{i_j})\leq e_i$ (since the Hamming metric is graphic). The word $W$ is built in the following way. For every index $i_j$  there are $K$ copies of the form {$(r_{i_j},\BoldData_i)$}, i.e., we move all the copies of each strand in both codewords to a word in the middle. 
It is easy to verify that {$W\in B^K_{(1,e_i,0)}(Z_1)\cap B^K_{(1,e_i,0)}(Z_2)$}, which is a contradiction since $\cC$ is a { {$(1,e_i,0)_K$}-\NameCodes}.

For the opposite direction, let {$Z_1,Z_2\in \cC_U $} be such that {$Z_1\neq Z_2 $}. We need to show that
 {$B^K_{(1,e_i,0)}(Z_1) \cap B^K_{(1,e_i,0)}(Z_2)=\emptyset  $}. From the assumption that $\cD(Z_1,Z_2)>2e_i$ and the definition of $\cD$, we have that there exists a data-field $\BoldData\in S(Z_1)$ such that there is no $\pi\in\bijection\left((I(\BoldData,Z_1),I(\BoldData,Z_2))\right)$ with $w_H(\pi_i)\leq2e_i$. 
 Equivalently, if we construct a bipartite graph $G=(L\cup R,E)$ where $L=I(\BoldData,Z_1), R=I(\BoldData,Z_2)$ and $E=\{(i,j)|i\in L, j\in R, d_H(i,j)\leq 2e_i\}$ then from Hall's marriage theorem  there is a subset $Y\subseteq L$ such that $|Y|>|N_G(Y)|$. 

We say that a read $s=(ind,\BoldData)$ is in the \emph{$e_i$ area of $Y$} if its index-field is at distance at most $e_i$ from at least one of the indices in $Y$. 
 Consider a general output word $w_1\in B^K_{(1,e_i,0)}(Z_1)$, the number of reads in $w_1$ that are in the $e_i$ area of $Y$  is at-least $K\cdot|Y|$. 
On the other hand, for every $w_2\in B^K_{(1,e_i,0)}(Z_2)$, the number of reads in $w_2 $ that are in the $e_i$ area of $Y$ is at most $K\cdot |N_G(Y)|<K\cdot|Y|$. Thus, since for every general output $w_1\in B^K_{(1,e_i,0)}(Z_1)$ the number of reads in the $e_i$ area of $Y$ is larger then the number of reads in the $e_i$ area of $Y$ for every general output $w_2\in B^K_{(1,e_i,0)}(Z_2)$, we have that {$B^K_{(1,e_i,0)}(Z_1) \cap B^K_{(1,e_i,0)}(Z_2)=\emptyset  $}.  
\end{IEEEproof}

Next, we study the case of $\frac{1}{2}+\frac{K \bmod{2}}{2K}\leq\tau<1$ (equivalent to $\tau<1$ such that $\frac{K}{2}\leq\lfloor{\tau K}\rfloor$) and present a similar necessary condition for this case. 

\begin{restatable}{lemma}{NeceseryTauSmallerThenOne}\label{lem:nec_cond}
For $\frac{1}{2}+\frac{K \bmod{2}}{2K}\leq\tau<1$ and $U\in MS({\cX}_{M,L,\nodollarindlen})$, if {$\cC_U$}  is a { {$(\tau,e_i,0)_K$}-\NameCodes} then $\cD(\cC_U)>e_i$.
\end{restatable}

\begin{IEEEproof}
Similar to the proof of the necessary condition in Theorem \ref{theorem-tau=1}, let $U=\{\!\{\BoldData_1,\BoldData_2,\dots,\BoldData_M\}\!\}\in MS(\cX_{M,L,\nodollarindlen})$ and assume that {$\cC_U\subseteq {\cX}_{M,L,\nodollarindlen}$} is a { {$(\tau,e_i,0)_K$}-\NameCodes}. Assume to the contrary that there are two codewords {$ Z_1,Z_2\in \cC_U$} such that {$\cD(Z_1,Z_2)\leq e_i$}. It will be shown that there exists {$W\in B^K_{(\tau,e_i,0)}(Z_1)\cap B^K_{(\tau,e_i,0)}(Z_2)$}. For every data-field {${\BoldData_i\in S(Z_1)}\hspace{0.04EX}$}, there exists $\pi_i\in\bijection\left((I(\BoldData_i,Z_1),I(\BoldData_i,Z_2))\right)$ such that $w_H(\pi_i)\leq e_i$. Thus, for every index $i_j\in I(\BoldData_i,Z_1)$ there exists a matched index $\pi_i(i_j)\in I(\BoldData_i,Z_2)$ with $d_H(i_j,\pi_i(i_j))\leq e_i$. The word $W$ is built in the following way. For every index $i_j$  there are $\lfloor \frac{K}{2}  \rfloor$ copies of the from {$({i_j},\BoldData_i)$} and $\lceil \frac{K}{2} \rceil$ copies of the from $(\pi_i({i_j}),\BoldData_i)$,  i.e., we move $\lceil\frac{K}{2}\rceil$ copies of each strand in $Z_1$ to its matched strand index in $Z_2$.
Since $\frac{K}{2}\leq\lfloor\tau K\rfloor$ it holds that $\lceil\frac{K}{2}\rceil\leq\lfloor\tau K\rfloor$ and hence $W\in B^K_{(\tau,e_i,0)}(Z_1)$. In addition $W$ can be obtained by taking every strand in $Z_2$ of the form $(\pi_i({i_j}),\BoldData_i)$ and move $\lfloor \frac{K}{2}  \rfloor$ copies to $({i_j},\BoldData_i)$.
Hence {$W\in B^K_{(\tau,e_i,0)}(Z_1)\cap B^K_{(\tau,e_i,0)}(Z_2)$}, which is a contradiction since $\cC$ is a { {$(\tau,e_i,0)_K$}-\NameCodes}
\end{IEEEproof}

Unfortunately, the opposite direction of Lemma~\ref{lem:nec_cond} does not generally hold, as seen in the following example.
\begin{example}
Consider the case of $K=2,e_i=1$, and $\tau=\frac{1}{2}$. Take $Z_1=\{(1100,\BoldData_1), (1000,\BoldData_1),$ $(0001,\BoldData_1)\}$ and $Z_2=\{(1100,\BoldData_1),(0001,\BoldData_1),(0101,\BoldData_1)\}$, it is easy to verify that $\cD(Z_1,Z_2)>1$. Nonetheless the following set   
$$W=\{(1000,\BoldData_1),(1001,\BoldData_1), (1100,\BoldData_1), (1101,\BoldData_1), (0001,\BoldData_1), (0101,\BoldData_1)\},$$
holds that  $W\in B_{(\frac{1}{2},1,0)}^2(Z_1)\cap B_{(\frac{1}{2},1,0)}^2(Z_2)$.

\end{example}
    
However, the opposite direction of Lemma~\ref{lem:nec_cond} holds if one assures that all data-fields in the stored sets differ. Let  $\overline{\cX}_{M,L,\nodollarindlen}$ denote all such sets, i.e., $${\overline{\cX}_{M,L,\nodollarindlen}=\{Z\in \cX_{M,L,\nodollarindlen}| \ |S(Z)|=M\}}.$$ Note that the size of $\overline{\cX}_{M,L,\nodollarindlen}$ is ${2^\nodollarindlen \choose M} {2^{L-\nodollarindlen} \choose M}M! $,  
and that $\overline{\cX}_{M,L,\nodollarindlen}\neq \emptyset$ if and only if $L-\nodollarindlen\geq \log_2(M)$. Although restricting to only sets in $\overline{\cX}_{M,L,\nodollarindlen}$ might reduce the number of information bits that is possible to store in the DNA storage system, it is verified in the next lemma, using the results from~\cite{SYLWZ20}, that for practical values of $\beta$, the cost of this constraint is at most a single redundancy bit. 

\begin{restatable}{lemma}{redAnalizeBar}
\label{lemma:redAna-DistinctData}
For $\beta<\frac{1}{2}\left(1-\frac{\nodollarindlen}{L}\right)$ it holds that $r(\overline{\cX}_{M,L,\nodollarindlen})<1.$ 
Furthermore, for $L$ that satisfies the following equation~\ref{eq:LforXBar} 
\begin{align}
L\geq \log(M)\big(3+2\log\log(M)\big)+8, \label{eq:LforXBar}  
\end{align}
there exists an efficient construction of $\overline{\cX}_{M,L,\log(M)}$ that uses a single redundancy bit. 
\end{restatable}
\begin{IEEEproof}
Denote $L'=L-\nodollarindlen$, we have that
\begin{align*}
r(\overline{\cX}_{M,L,\nodollarindlen})&=\log(|{\cX}_{M,L,\nodollarindlen}|)-\log\bigl(|\overline{\cX}_{M,L,\nodollarindlen}|\bigr)=\log\left({2^\nodollarindlen \choose M}{2^{(L-\nodollarindlen)M}}\right)-\log\left({2^\nodollarindlen \choose M} {2^{L-\nodollarindlen} \choose M}M!\right)\\
&=\log(2^{(L-\nodollarindlen)M})-\log\left({2^{L-\nodollarindlen} \choose M}M!\right)=\log(2^{L'M})-\log(\frac{2^{L'}!}{(2^{L'}-M)!})\\
&=\log \left(\frac{2^{L'}\cdot 2^{L'}\cdots 2^{L'}}{2^{L'}(2^{L'}-1)\cdots (2^{L'}-M+1)}\right)=\log \left(\prod_{i=0}^{M-1}\frac{2^{L'}}{2^{L'}-i}\right)\\&=-\log \left(\prod_{i=0}^{M-1}(1-\frac{i}{2^{L'}}\right).
\end{align*}

Using the inequality $(1-a)(1-b)>1-(a+b)$ for $a,b>0$, we get that $$\prod_{i=0}^{M-1}(1-\frac{i}{2^{L'}})>1-\sum_{i=0}^{M-1}\frac{i}{2^{L'}}=1-\frac{M(M-1)}{2^{L'+1}}>1-\frac{M^2}{2^{L'+1}}.$$ Thus, since $-\log$ is a monotonic decreasing function, it is derived that $r(\overline{\cX}_{M,L,\nodollarindlen})<-\log (1-\frac{M^2}{2^{L'+1}}).$

Lastly, since $\beta<\frac{1}{2}\left(1-\frac{\nodollarindlen}{L}\right)$ we have that $\frac{M^2}{2^{L'+1}}=\frac{2^{2\beta L}}{2^{L-\nodollarindlen+1}}=2^{L(2\beta-1)+\nodollarindlen-1}<\frac{1}{2}$. Thus, we conclude that $r(\overline{\cX}_{M,L,\nodollarindlen})<-\log (1-\frac{M^2}{2^{L'+1}})<-\log(1-\frac{1}{2})=1$.\\

Regarding the construction, in \cite{SYLWZ20}, the authors present such a construction(\cite{SYLWZ20}, Theorem 19) in the case that
$$L-2\log(M)\geq t+\log(M)+3.$$
Where $t\leq 5+2\log(M)\cdot\log(\log(M))$(\cite{SYLWZ20}, Lemma 24). Hence, if $$L\geq \log(M)\big(3+2\log\log(M)\big)+8,$$
then we have an efficient construction of $\overline{\cX}_{M,L,\log(M)}$.
\end{IEEEproof}

The notation $MS(\overline{\cX}_{M,L,\nodollarindlen})$ is used to denote the set of all possible data-field multisets of elements in $\overline{\cX}_{M,L,\nodollarindlen}$, which are in essence sets. The next lemma presents a sufficient condition for such sets.

\begin{restatable}    
{lemma}{SufficetTauSmallerThenOne}\label{lem:suffConTau>half}
 For $\frac{1}{2}+\frac{K \bmod{2}}{2K}\leq\tau<1$ and $U\in MS(\overline{\cX}_{M,L,\nodollarindlen})$,  if  {$\cD(\cC_U)>e_i$} then $\cC_U$ is a {$(\tau,e_i,0)_K$}-\NameCodes.
\end{restatable}
\begin{IEEEproof}
The proof is similar to the proof of the sufficient condition in Theorem \ref{theorem-tau=1}. Let {$Z_1,Z_2\in \cC_U $} be such that {$Z_1\neq Z_2 $}. We need to show that
 {$B^K_{(\tau,e_i,0)}(Z_1) \cap B^K_{(\tau,e_i,0)}(Z_2)=\emptyset  $}. From the assumption that $\cD(Z_1,Z_2)>e_i$ and the definition of $\cD$, we have that there exists a data-field $\BoldData\in S(Z_1)$ such that $d_H(I(\BoldData,Z_1),I(\BoldData,Z_2))> e_i$. Since $\tau<1$ in every general output $w_1\in B^K_{(\tau,e_i,0)}(Z_1)$ there is at least one read of the form $(I(\BoldData,Z_1),\BoldData)$, on the other hand since $d_H(I(\BoldData,Z_1),I(\BoldData,Z_2))> e_i$ in every general output $w_2\in B^K_{(\tau,e_i,0)}(Z_2)$ there are zero reads of the form $(I(\BoldData,Z_1),\BoldData)$. Hence, $B^K_{(\tau,e_i,0)}(Z_1) \cap B^K_{(\tau,e_i,0)}(Z_2)=\emptyset$.
\end{IEEEproof}

The next corollary summarizes this discussion. 

\begin{corollary}\label{col-tau<1}
For $\frac{1}{2}+\frac{K \bmod{2}}{2K}\leq\tau<1$ and $U\in MS(\overline{\cX}_{M,L,\nodollarindlen})$, $\cC_U$ is a {$(\tau,e_i,0)_K$}-\NameCodes\ if and only if $\cD(\cC_U)>e_i$. 
\end{corollary} 

We continue to study the case of $\tau<\frac{1}{2}+\frac{K \bmod{2}}{2K}$ in Lemma~\ref{lem:tau<halfBar} (recall that this condition implies that $\lfloor{\tau K}\rfloor<\frac{K}{2}$).

\begin{restatable}{lemma}{BarForTauLessHalf}\label{lem:tau<halfBar}
For $\tau<\frac{1}{2}+\frac{K \bmod{2}}{2K}$, it holds that for every $e_i$, $\overline{\cX}_{M,L,\nodollarindlen}$ is a {$(\tau,e_i,0)_K$}-\NameCodes.
\end{restatable}

\begin{IEEEproof}
In this case, the output strands enjoy the property which is referred to at \cite{SYLWZ20} as the dominance property, i.e., if the strands are clustered by their index, and every such cluster is partitioned into subsets based on the original index part, the largest subset will be the correct subset. 
This is indeed the situation, as the data part of every strand is different and free of errors, thus, the correct subset would be of size at least {$\lfloor \frac{K}{2} \rfloor+1$} while all others subsets would be of size at most {$\lfloor \frac{K}{2} \rfloor$}. Hence the naive algorithm which clusters the strands by their index and matches every index with the data that fits with the majority of this cluster, retrieves the original input successfully.
\end{IEEEproof}

\subsection{Codes for a Fixed Data-Field Set}\label{subsec:index_cor}
So far in the paper we focused on properties and conditions of DNA-correcting codes that guarantee successful decoding of the data. In particular, Corollary~\ref{corollary-codeForDifferentMultiSets} showed that it is enough to construct codes for every data-field multiset $U\in MS({\cX}_{M,L,\nodollarindlen})$ independently, while the conditions concerning the DNA-distance were established in Theorem \ref{theorem-tau=1}, and Lemmas \ref{lem:nec_cond}, \ref{lem:suffConTau>half} and \ref{lem:tau<halfBar}. These conditions depend upon the value of $\tau$ and whether $U$ is a set or a multiset. In Lemma~\ref{lemma:redAna-DistinctData}, it was shown that for all practical values of $\beta$, restricting to using only sets in $\overline{\cX}_{M,L,\nodollarindlen}$ imposes only a single bit of redundancy and therefore, the next section provides DNA-correcting codes for $\overline{\cX}_{M,L,\nodollarindlen}$ when $e_d=0$.

Note that for $U,U'\in MS(\overline{\cX}_{M,L,\nodollarindlen})$ it holds that $A_{M,L,\nodollarindlen}(\tau,e_i,0,K)_U=A_{M,L,\nodollarindlen}(\tau,e_i,0,K)_{U'}$ and thus for the rest of the paper we fix $U=\{\BoldData_1,\BoldData_2,\dots,\BoldData_M\}\in MS(\overline{\cX}_{M,L,\nodollarindlen})$ and our goal is to find \NameCodes s for $U$ with a given DNA-distance. Note that for $Z = \{(ind_1,\bfu_1),\dots,(ind_{M},\bfu_{M})\}, Z' = \{(ind_1',\bfu_1),\dots,(ind_{M}',\bfu_{M})\}$, it holds that 
$$\cD(Z_1,Z_2) = \max_{1\leq i\leq M}d_H(ind_i,ind_i').$$

Thus, we focus on studying the next family of codes for the index-fields.
\begin{definition}\label{def:index codes}
Let $I(\nodollarindlen,M)=\{(ind_1,\dots ,ind_M)| \forall i: ind_i\in \{{0,1}\}^{\nodollarindlen},\forall i\neq j: ind_i\neq ind_j \}$. For every two codewords $\bfc = (c_1,\ldots,c_M),\bfc'=(c'_1,\ldots,c'_M)\in I(\nodollarindlen,M)$, their \emph{index-distance} is defined by $\cD_I(\bfc,\bfc') \triangleq \max\limits_{1\leq i\leq M}d_H(c_i,c_i')$ and for a code $\cC \subseteq I(\nodollarindlen,M)$, its \emph{index-distance} is defined by $\cD_I(\cC) \triangleq \min\limits_{\bfc\neq \bfc'\in \cC} \cD_I(\bfc,\bfc')$. A  code $\cC \subseteq I(\nodollarindlen,M)$ will be called an {\emph{\indexCode}} if $\cD_I(\cC)\geq d$. We denote by $F(\nodollarindlen,M,d)$ the size of a maximal \indexCode. 
\end{definition}

\begin{example}\label{Ex:matrix}
The rows of the following matrix form a $(2,4,2)$ index-correcting code, while each row corresponds to a codeword,
\begin{align}    
P=\begin{pmatrix}
00 & 01 & 11 & 10\\
00 & 11 & 10 & 01\\
00 & 10 & 01 & 11\\
11 & 01 & 00 & 10\\
11 & 00 & 10 & 01\\
11 & 10 & 01 & 00
\end{pmatrix}.
\end{align}
One can verify, that for every two different rows $i,i'$ in $P$, there exists a column $j$ such that $d_H(P(i,j),P(i',j))=2$.
\end{example}

The motivation for studying this family of codes comes from the following observation which results from Theorem \ref{theorem-tau=1}, Corollary \ref{col-tau<1}, and Lemma \ref{lem:tau<halfBar}.

\begin{observation}
For $U\in MS(\overline{\cX}_{M,L,\nodollarindlen})$, it holds that
\begin{small}  
$$A_{M,L,\nodollarindlen}(\tau,e_i,0,K)_U=\begin{cases}
F(\nodollarindlen,M,2e_i+1),   & \tau=1 \\
F(\nodollarindlen,M,e_i+1),   & \frac{K}{2}\leq\lfloor{\tau K}\rfloor<K \\
 {2^{\nodollarindlen} \choose M}M!,   & \lfloor{\tau K}\rfloor<\frac{K}{2}.
\end{cases}$$
\end{small}
\end{observation}

Note that the study of index-correcting codes and in particular the value of $F(\nodollarindlen,M,d)$ is interesting on its own and can be useful for other problems, independently of the problem of designing codes for DNA storage. The next section is dedicated to a careful investigation of these codes. 

\section{Index-Correcting Codes}\label{Sec:constructionSection}
In this section, we study index-correcting codes. We start with the special case of $\nodollarindlen=\log(M)$.

\subsection{$\nodollarindlen=\log(M)$}
In this case, every possible codeword in $I(\log(M),M)$ is a permutation over $\{0,1\}^{\log(M)}$, and for $f,g\in I(\log(M),M)$, their index-distance is equivalent to the $\ell_\infty$ distance over the Hamming distance of the indices, i.e., $$\cD_I(f,g)=\max\limits_{i\in \{0,1\}^{\log(M) }} d_H(f(i),g(i)).$$     
For $f\in I(\log(M),M)$, let $B_r(f)$ be the ball of radius $r$ centered at~$f$ in $I(\log(M),M)$, i.e., 
$$B_r(f)=\{g\in I(\log(M),M)| \cD_I(f,g)\leq r\}.$$
In this case, it holds that  $\cD_I$ is right invariant, i.e., for $f,g,p\in I(\log(M),M)$, we have that $\cD_I(f,g)=\cD_I(f\circ p,g\circ p)$, and thus the size of the balls in $I(\log(M),M)$ is the same. Let $B_{r,M}$ denote the size of the balls of radius $r$ in $I(\log(M),M)$. An important matrix with respect to $B_{r,M}$ is the matrix  $A_{r,M}=(a_{i,j}) $ of size $M\times M$ which is defined by 
$$a_{i,j}={\begin{cases}
1 & d_H(i,j)\leq r\\ 
0 & \text{otherwise}    
\end{cases}}, i,j\in \{0,1\}^{\log(M)}.$$
Let $per (A)$ denote the permanent of a square matrix $A$. For completeness, the definition of a permanent of a square matrix is presented next.

\begin{definition}
 Let $S_n$ denote all permutations over $[n]$. For an $n\times n$ matrix $A=(a_{i,j})$, the permanent of $A$ is defined as $per(A)\triangleq \sum\limits_{f\in S_n}\prod\limits_{i\in [n]}a_{i,f(i)} $.   
\end{definition}

The next lemma follows in a similar way to the one presented in~\cite{SO17}. 
\begin{restatable}{lemma}{ballsEqualPer}\label{lem:Ball=Perm}
It holds that $B_{r,M}=per (A_{r,M})$.
\end{restatable}
\begin{IEEEproof}
Let $ID_M$ be the identity permutation over $\{0,1\}^{\log(M)}$. Then $B_{r,M}=B_{r}(ID_M)$ and from the definition of the permanent we have that:
 \begin{align*}
  per(A_{r,M}) & =  \sum_{f\in S_M}\prod_{i\in \{0,1\}^{\log (M)}}a_{i,f(i)}\\
  & \overset{\mathrm{(a)}}{=}|\{f\in S_M) |\forall i\in  \{0,1\}^{\log(M)}: d_H(i,f(i))\leq r\}|\\
  & = |\{f\in S_M) |\cD_I(ID_M,f)\leq r\}|=|B_{r}(ID_M)|=B_{r,M},
 \end{align*} 
where (a) holds because of the shape of the matrix $A_{r,M}$.
\end{IEEEproof}

Next, two bounds on $F(\log(M),M,d)$ are presented. Lemma~\ref{lem:SpherPacking} uses the sphere packing bound with a known bound on the permanent of a matrix, while Lemma~\ref{lem:Singelton} uses a method that is similar to the proof of the Singleton bound.

\begin{restatable}{lemma}{SpherPackingBound}\label{lem:SpherPacking}
 It holds that $$F(\log(M),M,d)\leq \frac{M^M}{({\sum_{i=0}^{\lfloor \frac{d-1}{2}\rfloor} {\log(M) \choose i}})^M}.$$
\end{restatable} 
\begin{IEEEproof}
By the Van der Waerden conjecture, which was proved in \cite{F81}, it is known that for a binary $M\times M$ matrix for which the sum of the elements in any row or column is $k$, the permanent of the matrix is lower bounded by  $\frac{M!k^M}{M^M}$. Note that the number of ones in every row and column of  $A_{\lfloor\frac{d-1}{2}\rfloor,M}$ is exactly $\sum_{i=0}^{\lfloor \frac{d-1}{2}\rfloor} {\log(M) \choose i}$, thus, since the balls of radius $\lfloor \frac{d-1}{2}\rfloor$ are mutually disjoint we have that 
\begin{align*}   
&F(\log(M),M,d)\frac{M!({\sum_{i=0}^{\lfloor \frac{d-1}{2}\rfloor} {\log(M) \choose i}})^M}{M^M}\leq F(\log(M),M,d)\cdot B_{\lfloor\frac{d-1}{2}\rfloor,M}\leq M!.
\end{align*}
 Thus, achieving the statement of the lemma.
\end{IEEEproof}

\begin{restatable}{lemma}{SingeltongBound}\label{lem:Singelton}

 $F(\log(M),M,d)\leq\frac{M!}{(2^{d-1} !)^{2^{\log(M)-d+1}}}$. 
\end{restatable}

\begin{IEEEproof}
Let  $\cC $ be  a \indexLogMCode\ and let $P$ be the matrix whose rows are the codewords of $\cC$. Given a row of indices in $P$, $ind_1,ind_2,\ldots,ind_M$, we divide every such $ind_j$ to two parts, $a_j\in \{0,1\}^{\nodollarindlen-d+1}$ which is the first $\log(M)-d+1$ bits of $ind_j$ and $b_j\in \{0,1\}^{d-1}$ which is the last $d-1$ bits of $ind_j$. We concatenate all $a_j$ to form a vector ${\bfa}\in \left(\{{0,1\}^{\log(M)-d+1}}\right)^M$. Note that every such vector $\bfa$ can only appear once in $P$ (as if there are two different rows $i,i'$ that form this vector, then the Hamming distance between indices of the same column is at most $d-1$, which implies that the index-distance between those two rows is at most $d-1<d$). Hence, the number of possible rows in $P$, is bounded from above by the number of possible vectors $\bfa\in \left(\{{0,1\}^{\log(M)-d+1}}\right)^M$. Since all the $\{0,1\}^{\log(M)}$ indices must appear in every row, every vector in $\{0,1\}^{\log(M)-d+1}$ must appear $2^{d-1}$ times in $\bfa$. Thus, from the multinomial theorem the number of possible vectors $\bfa$ is $\frac{M!}{(2^{d-1} !)^{2^{\log(M)-d+1}}}$.
Since we started with a general \indexLogMCode, we conclude that $F(\log(M),M,d)\leq\frac{M!}{(2^{d-1} !)^{2^{\log(M)-d+1}}}$.    
\end{IEEEproof}

We obtain the following corollary by assigning $d=2$ to Lemma~\ref{lem:Singelton}.
\begin{corollary}\label{CorBoundFord=2}
 $F(\log(M),M,2)\leq \frac{M!}{2^{\frac{M}{2}}}.$   
\end{corollary}
Note that the code in Example \ref{Ex:matrix} achieves the bound in Corollary~\ref{CorBoundFord=2} for $M=4$, and hence the bound of Lemma~\ref{lem:Singelton}  can be tight in some cases.

Next, we present a construction by building a matrix whose rows form a \indexLogMCode. 
Such a matrix whose rows form an \indexCode\ will be called an \emph{\MatrixCorrecting}.
The construction uses codes over $\{0,1\}^{\log(M)}$ with Hamming distance $d$ and afterward an example for small values of $d$ is presented.

\begin{construction}\label{ Construction:Log(M)}
Let there be an optimal, with respect to the Hamming distance, binary linear code $\cC$ of length $\log(M)$ with minimum distance $d$. Denote by $A$ the size of $\cC$ and note that the $\frac{M}{A}$ cosets of $\cC$ form a partition of $\{0,1\}^{\log(M)}$. Denote the cosets of $\cC$ by $\cC_1=\cC,\cC_2,\dots,\cC_{\frac{M}{A}}$. We start by building a matrix that consists of $(A!)^\frac{M}{A}$ rows, where the first $A$ entries of every row are permutations over the first coset, the second $A$ entries are permutations over the second coset, and so on. Since the entries of every column belong to the same coset, the distance between different rows is at least  $d$.
Next, we take every coset $\cC_i$, for $2\leq i\leq \frac{M}{A}$, and remove from it all words that are at distance at most $d$ from the zero vector $\mathbf{0}$, denote by $\cC'_i$ the achieved codes and let $\cC'=\bigcup_{i=2}^{\frac{M}{A}}\cC'_i$. Then, for every $c'\in \cC'$, we can look at all the rows in the matrix where $c'$ and $\mathbf{0}$ are fixed to the first entry of their coset (note that there are $(A-1)!^{2} (A!)^{\frac{M}{A}-2}$ such rows) and add the same row where we replace the entry of $c'$ with $\mathbf{0}$,  adding in total  $(A-1)!^2|\cC'| (A!)^{\frac{M}{A}-2}$ more rows. In conclusion, we achieve a code with $(A!)^\frac{M}{A}+(A-1)!^2|\cC'| (A!)^{\frac{M}{A}-2}=(A!)^\frac{M}{A}\left(1+|\cC'|\cdot\frac{1}{A^2}\right)$ codewords.

\end{construction}
We show how to apply Construction~\ref{ Construction:Log(M)} in the next example.\begin{restatable}{example}{ExConstructionForLog}\label{ExConstruc}
 We apply Construction~\ref{ Construction:Log(M)} to the case of $d=2$ and $d=3$. For $d=2$ we have that the maximal linear code $\cC$ is the parity code with $|\cC|=\frac{M}{2}$, and that $|\cC'|=\frac{M}{2}-\log(M)$. 
 Thus in this case we get a  $(\log(M),M,2)$ index-correcting code with size of $(\frac{M}{2}!)^2(1+\frac{4(\frac{M}{2}-\log(M))}{M^2})$.
 
 For $d=3$ and $\log(M)=\nodollarindlen=2^m-1$ we have that the maximal linear code $\cC$ is the binary Hamming code with $|\cC|=2^{2^m-1-m}$, and that there are $2^m$ cosets (including the code itself). In addition, every coset $\cC_i\neq \cC$ has one word of weight $1$ and $\frac{2^m-2}{2}$ words of  weight $2$, thus $|\cC'_i|=2^{2^m-1-m}-\frac{2^m-4}{2}$. Hence, by applying Construction~\ref{ Construction:Log(M)} we get a $(2^m-1,2^{2^m-1},3)$ index-correcting code of size $(2^{2^m-1-m}!)^{2^m}(1+(2^m-1)(2^{2^m-1-m}-\frac{2^m-4}{2})\frac{1}{(2^{2^m-1-m})^2})$.
Using that $m=\log(\nodollarindlen+1)=\log(\log(M)+1)$ we have that the size of the code is
\begin{align*}
\left(\frac{M}{\log(M)+1}!\right)&^{\log(M)+1}\cdot\left(1+\log(M)\left(\frac{M}{\log(M)+1}-\frac{\log(M)-3}{2}\right)\frac{(\log(M)+1)^2}{M^2}\right)\\
&=\left(\frac{M}{\log(M)+1}!\right)^{\log(M)+1}\left(1+\Theta\left (\frac{\log^2(M)}{M}\right)\right).
\end{align*}

Hence, we have a $(2^m-1,2^{2^m-1},3)$ index-correcting code of size $(\frac{M}{\log(M)+1}!)^{\log(M)+1}\left(1+ g(M)\right)
$, where $g(M)=\Theta(\frac{\log^2(M)}{M})$. 

\end{restatable}

\subsection{$\nodollarindlen>\log(M)$}

In this case, the set of possible indices is larger than the number of strands. Next we show how to construct an  $(\nodollarindlen',M,d)$ index-correcting code from an \indexCode\ for $\nodollarindlen<\nodollarindlen'$. 

\begin{restatable}{lemma}{ProofConsForLargel}\label{ConstructionForLargel}
For $\nodollarindlen'=\nodollarindlen+\lceil \frac{d}{2}\rceil$ it holds that $F(\nodollarindlen',M,d)\geq F(\nodollarindlen,M,d)\cdot 2^M.$  
\end{restatable}

\begin{IEEEproof}
We show a construction of a matrix $P'$ using an optimal \MatrixCorrecting\  $P$ with $F(\nodollarindlen,M,d)$ rows, and later we prove that the matrix $P'$ is a legal \lMatCorrecting\ with $F(\nodollarindlen,M,d)\cdot2^M$ rows. The construction is specified next iteratively.
\begin{enumerate}
    \item Obtain a matrix $P'_0$ by adding $\lceil \frac{d}{2}\rceil$ bits of $0$ at the end of every index in $P$. 
    \item For $j=1,2,\ldots, M$: 
        For every row $i$ of $P'_{j-1}$, add a similar row of indices $b_j(i)$, which differs from the $i$-th row only in the $j$-th column (i.e., the $j$-th index of the row). The difference is that in $b_j(i)$, the first $\lceil \frac{d}{2}\rceil$ bits and last $\lceil \frac{d}{2}\rceil$ bits of the $j$-th index, are the complement of the corresponding index in row $i$. We denote the matrix obtained after the $j$-th step by $P'_{j}$.
\end{enumerate}

We are left to show that the above construction ends with a legal \lMatCorrecting\ with $F(\nodollarindlen,M,t)\times2^M$ rows. First, since for every step $j$ the number of rows is multiplied by two, we have at the end a matrix $P'=P'_M$ with $F(\nodollarindlen,M,t)\times2^M$ rows. 

Next, we show by induction that after every step $j$, $P'_j$ is an \lMatCorrecting. The base case is simple, as $P$ is an \MatrixCorrecting\ and hence $P'_0$ is an \lMatCorrecting. 
For the induction step, on step $j$ we start with a matrix $P'_{j-1}$ which is an \lMatCorrecting. First, note, that the indices in every row in $P'_j$ are disjoint, as otherwise we can complement the last $\lceil \frac{d}{2}\rceil$ and the first $\lceil \frac{d}{2}\rceil$ bits of these indices and have that there was a row in $P'_0$ which had two identical indices. Now, take two general rows $i,i'\in  P'_{j-1}$, the distance between rows $b_j(i)$ and $b_j(i')$ is the same as the distance between rows $i$ and $i'$, and the distance between rows $i$ and $b_j(i)$ is $\lceil \frac{d}{2}\rceil+\lceil \frac{d}{2}\rceil\geq d$ (as they differ in the first and last $\lceil \frac{d}{2}\rceil$ components in index $j$). Thus, we are left to check the distance condition for rows $i$ and $b_j(i')$.
But the distance between rows $i$ and $b_j(i')$ can only increase from the distance between rows $i$ and $i'$ (because they will differ at the last $\lceil\frac{d}{2}\rceil$ bits) and $P_j'(i)$ is an \lMatCorrecting. Hence, the distance property holds and $P'_j$ is an \lMatCorrecting.
\end{IEEEproof}

\begin{example}\label{ExConsWithSteps}
The next matrices are obtained from the matrix in Example \ref{Ex:matrix} after step $j=1$ and step $j=2$.
\begin{small}
\begin{align}    
P'_1=\begin{pmatrix}
000 & 010 & 110 & 100\\
000 & 110 & 100 & 010\\
000 & 100 & 010 & 110\\
110 & 010 & 000 & 100\\
110 & 000 & 100 & 010\\
110 & 100 & 010 & 000\\
101 & 010 & 110 & 100\\
101 & 110 & 100 & 010\\
101 & 100 & 010 & 110\\
011 & 010 & 000 & 100\\
011 & 000 & 100 & 010\\
011 & 100 & 010 & 000
\end{pmatrix}
\end{align}
\begin{align}    
P'_2=\begin{pmatrix}
000 & 010 & 110 & 100\\
000 & 110 & 100 & 010\\
000 & 100 & 010 & 110\\
110 & 010 & 000 & 100\\
110 & 000 & 100 & 010\\
110 & 100 & 010 & 000\\
101 & 010 & 110 & 100\\
101 & 110 & 100 & 010\\
101 & 100 & 010 & 110\\
011 & 010 & 000 & 100\\
011 & 000 & 100 & 010\\
011 & 100 & 010 & 000\\
000 & 111 & 110 & 100\\
000 & 011 & 100 & 010\\
000 & 001 & 010 & 110\\
110 & 111 & 000 & 100\\
110 & 101 & 100 & 010\\
110 & 001 & 010 & 000\\
101 & 111 & 110 & 100\\
101 & 011 & 100 & 010\\
101 & 001 & 010 & 110\\
011 & 111 & 000 & 100\\
011 & 101 & 100 & 010\\
011 & 001 & 010 & 000
\end{pmatrix}
\end{align}
\end{small}

\end{example}
This example leads to the following recursive construction.

\begin{lemma}\label{lem:recurConFord=2}
It holds that
$$F(\log(M)+1,2M,2)\geq 2^M F(\log(M),M,2)^2.$$    
\end{lemma}
\begin{IEEEproof}
Let $P$ be an $(\log(M),M,2)$ matrix with $F(\log(M),M,2)$ rows. We use the construction specified in Lemma \ref{ConstructionForLargel} to obtain a $(\log(M)+1,M,2)$ matrix $P'$ with $2^M\cdot F(\log(M),M,2)$ rows. 
In order to convert the matrix $P'$ into a $(\log(M)+1,2M,2)$ matrix, $M$ missing indices need to be added to every row in $P'$. Note that for each row $i$ in $P'$, we can add $F(\log(M),M,2)$ similar rows that have the same first $M$ indices as in row $i$, and the last $M$ indices obtain the distance property. This is because the worst case is when all the indices in $i$ finish with the same bit, and this case is 
equivalent to finding a maximal   $(\log(M),M,2)$ matrix. 
\end{IEEEproof}
\begin{example}
The next six rows can be obtained by adding the $F(2,4,2)$ rows of indices concatenated with a $1$ bit in the end, to the row $\begin{pmatrix}
000& 010 & 110 & 100
\end{pmatrix}$.
\begin{small}
\begin{align*}    
\begin{pmatrix}
000 & 010 & 110 & 100 &001 & 011 & 111 & 101\\
000 & 010 & 110 & 100 &001 & 111 & 101 & 011\\
000 & 010 & 110 & 100 &001 & 101 & 011 & 111\\
000 & 010 & 110 & 100 &111 & 011 & 001 & 101\\
000 & 010 & 110 & 100 &111 & 001 & 101 & 011\\
000 & 010 & 110 & 100 &111 & 101 & 011 & 001
\end{pmatrix}.
\end{align*}
\end{small}

Note that if we delete the last bit of every index in the last $4$ columns, we obtain the matrix in Example~\ref{Ex:matrix}. The next six rows, show how to extend the row     $\begin{pmatrix}
011& 100 & 111 & 101
\end{pmatrix}$ into $6$ rows where all the length-$3$ binary vectors appear.
\begin{small}
\begin{align*}    
\begin{pmatrix}
011 & 100 & 111 & 101& 001 & 110 & 010 & 000\\
011 & 100 & 111 & 101& 001 & 010 & 000 & 110\\
011 & 100 & 111 & 101& 001 & 000 & 110 & 010\\
011 & 100 & 111 & 101& 010 & 110 & 001 & 000\\
011 & 100 & 111 & 101& 010 & 001 & 000 & 110\\
011 & 100 & 111 & 101& 010 & 000 & 110 & 100
\end{pmatrix}
\end{align*}
\end{small}
\end{example}

By applying Lemma~\ref{lem:recurConFord=2} recursively and the fact that  $F(2,4,2)=6$ (from Example \ref{Ex:matrix} and Lemma~\ref{lem:Singelton}), we obtain the following Corollary.

\begin{corollary}\label{Cor:RecForD=2}
For $M\geq4$ a power of two we have that $F(\log(M)+1,2M,2)\geq 2^{M(\log(M)-1)}\cdot 6^{(2^{\log(M)-1})}.$    
\end{corollary}

\section{Erroneous Data-Field}\label{Sec:ErrornesData}
\subsection{Sufficient and Necessary Conditions}
In this subsection, we study the case of $e_d>0$ and present conditions for a code to be a $(\tau,e_i,e_d)_K$-\NameCodes. For this purpose, we define a generalization to the DNA distance. For $Z_1,Z_2\in \cX_{M,L,\nodollarindlen}$, we say that $\cD(Z_1,Z_2)>(r,t)$ if for every $\pi\in BI(Z_1,Z_2)$, there exists a strand $s=(ind,\BoldData)\in Z_1$ such that $d_H(ind, \pi(ind))>r$ or $d_H(\BoldData,\pi(\BoldData))>t$. For a code $\cC\subseteq \cX_{M,L,\nodollarindlen}$, if for every $Z_1\neq Z_2\in\cC$ it holds that $\cD(Z_1,Z_2)>(r,t)$, then we say that $\cD(\cC)>(r,t)$. Note that when $t=0$ this definition is equivalent to the regular DNA-distance.

Theorem~\ref{ThoremEd>0tau=1}, Lemma~\ref{e_d>0And0.5<Tau<1} is a generalization of Theorem~\ref{theorem-tau=1}, Lemma \ref{lem:nec_cond}, respectively, and can be proved by similar arguments. We note that the proofs of Theorem~\ref{ThoremEd>0tau=1} and Lemma~\ref{e_d>0And0.5<Tau<1} were also obtained in a parallel work by Wu~\cite{HWU23}. For completeness of this paper, we present the proofs here as well.

\begin{restatable}{theorem}{edLargerAndTauEqualsOne}
\label{ThoremEd>0tau=1}
A code $\cC\subset \cX_{M,L,\nodollarindlen}$ is a $(1,e_i,e_d)_K$-\NameCodes\ if and only if $\cD(\cC)>(2e_i,2e_d)$.  
\end{restatable}

\begin{IEEEproof}
For the first direction, assume that $\cC$ is a $(1,e_i,e_d)_K$-\NameCodes\ and assume for contrary that there exist two different codewords of $\cC$, $Z_1\neq Z_2\in\cC$, and a matching $\pi'\in BI(Z_1,Z_2)$ such that for every strand $s=(ind_j,\BoldData_j)\in Z_1$ it holds that $d_H(ind_j,\pi'(ind_j))\leq2e_i$ and  $d_H(\BoldData_j,\pi'(\BoldData_j)\leq 2e_d$. As in the proof of Theorem~\ref{theorem-tau=1}, for every strand in $Z_1$ there is a matched strand in $Z_2$, such that we can move all the $K$ copies of the strands to a noisy strand in the middle (since $\tau=1$), this will end in a word  $W\in B^K_{(1,e_i,e_d)}(Z_1) \cap B^K_{(1,e_i,e_d)}(Z_2)$, which is a contradiction.

For the opposite direction, let $Z_1\neq Z_2\in \cC$. We need to show that $B^K_{(1,e_i,e_d)}(Z_1) \cap B^K_{(1,e_i,e_d)}(Z_2)=\emptyset$. Similarly to the proof of Theorem~\ref{theorem-tau=1}, if we construct a bipartite graph $G=(L\cup R,E)$, where $L=Z_1$, $R=Z_2$, and $E=\{\left((ind,\BoldData),(ind',\BoldData')\right)|(ind,\BoldData)\in Z_1,(ind',\BoldData')\in Z_2,d_H(ind,ind')\leq 2e_i, d_H(\BoldData,\BoldData')\leq 2e_d\}$, then from Hall's theorem there is a subset $Y\subseteq L$ such that $|Y|>|N_G(Y)|$. Hence, for every general output  $W_1\in B^K_{(1,e_i,e_d)}(Z_1)$ the number of reads in the $(e_i,e_d)$ area of $Y$ is at least $|Y|\cdot K$, whereas in every general output $W_2\in B^K_{(1,e_i,e_d)}(Z_2)$, the number of reads in the $(e_i,e_d)$ area of $Y$ is at most $|N_G(Y)|\cdot K<|Y|\cdot K$.
\end{IEEEproof}

\begin{restatable}{lemma}{GeneralizationNeccesaryCondition}\label{e_d>0And0.5<Tau<1}
Let $\cC\subset\cX_{M,L,\nodollarindlen}$ and $\frac{1}{2}+\frac{K\bmod{2}}{2K}\leq\tau<1$. If $\cC$ is a $(\tau,e_i,e_d)_K$-\NameCodes\ then
$\cD(\cC)>(e_i,e_d)$.  
\end{restatable}

\begin{IEEEproof}
Assume to the contrary that there are two codewords $Z_1\neq Z_2\in \cC$ and $\pi\in BI(Z_1,Z_2)$ such that for every strand $s=(ind_j,\BoldData_j)\in Z_1$ it holds that $d_H(ind_j,\pi(ind_j))\leq e_i$ and $d_H(\BoldData_j,\pi(\BoldData_j))\leq e_d$.
As in the proof of Lemma~\ref{lem:nec_cond}, we build a word $W\in B^K_{(\tau,e_i,e_d)}(Z_1) \cap B^K_{(\tau,e_i,e_d)}(Z_2)$ by moving $\lceil \frac{K}{2}\rceil$ copies of each strand in $Z_1$ to its matched strand in $Z_2$. Note that due to the condition on $\tau$, $W$ is indeed in $B^K_{(\tau,e_i,e_d)}(Z_1) \cap B^K_{(\tau,e_i,e_d)}(Z_2)$.
\end{IEEEproof}

As in the case for $e_d=0$, the opposite direction of Lemma \ref{e_d>0And0.5<Tau<1} is not necessarily true. To overcome this, the following subspace of ${\cX}_{M,L,\nodollarindlen}$ was defined in~\cite{HWU23},
\begin{align*}
\overline{\cX}_{M,L,\nodollarindlen}^{(r_1,r_2)}=\{&Z\in {\cX}_{M,L,\nodollarindlen}|\forall s_1=(ind_1,\BoldData_1), s_2=(ind_2,\BoldData_2)\in Z,  \textit{ such that } s_1\neq s_2,\\& d_H(ind_1,ind_2)>r_1\textit{ or } d_H(\BoldData_1,\BoldData_2)>r_2\}.
\end{align*}

Note that when $\nodollarindlen=\log(M)$, a code  $\cC\subseteq\overline{\cX}_{M,L,\log(M)}^{(r_1,r_2)}$ is refereed to in~\cite{SYLWZ20} as an $(r_1,r_2+1)$-clustering-correcting code (only to the data-field was added $+1$ because in~\cite{SYLWZ20} the set was defined with a weak inequality). 
The next lemma explains why we focus on the set $\overline{\cX}_{M,L,\nodollarindlen}^{(r_1,r_2)}$.

\begin{lemma}\label{Lem:Tau>0.5ConstruFirstCond}(\cite{HWU23}, Lemma 4)
Assume that $\tau<1$ such that $\frac{K}{2} \leq\lfloor\tau K\rfloor$, and let  $Z_1,Z_2\in \overline{\cX}_{M,L,\nodollarindlen}^{(2e_i,2e_d)}$, then $$B^K_{(\tau,e_i,e_d)}(Z_1) \cap B^K_{(1,e_i,e_d)}(Z_2)\neq\emptyset$$
if and only if $\cD(Z_1,Z_2)\leq (e_i,e_d)$.
\end{lemma} 
For $\frac{K}{2}\leq \lfloor\tau K\rfloor<\frac{MK}{2M-1}$, Wu has prove that restricting to $\overline{\cX}_{M,L,\nodollarindlen}^{(e_i,e_d)}$ is enough.

\begin{lemma}\label{Lem:E_d>0Tau>0.5BetterCon}(\cite{HWU23}, Lemma 5)
Assume that $\frac{K}{2}\leq \lfloor\tau K\rfloor<\frac{MK}{2M-1}$, and let  $Z_1,Z_2\in \overline{\cX}_{M,L,\nodollarindlen}^{(e_i,e_d)}$, then $$B^K_{(\tau,e_i,e_d)}(Z_1) \cap B^K_{(1,e_i,e_d)}(Z_2)\neq\emptyset$$
if and only if $\cD(Z_1,Z_2)\leq (e_i,e_d)$.    
\end{lemma}

Next, we study the case of $\tau<\frac{1}{2}+\frac{K \bmod{2}}{2K}$. The following lemma is the generalization of Lemma~\ref{lem:tau<halfBar} for $e_d>0 $ and is proved in a similar way. 
\begin{lemma}\label{Lem:XbarForE_d>0AndT<half}
For $\tau<\frac{1}{2}+\frac{K \bmod{2}}{2K}$, it holds that for every $e_i$ and $e_d$, $\overline{\cX}_{M,L,\nodollarindlen}^{(2e_i,2e_d)}$ is a {$(\tau,e_i,e_d)_K$}-\NameCodes.    
\end{lemma}

\begin{IEEEproof}
For every $Z\in \overline{\cX}_{M,L,\nodollarindlen}^{(2e_i,2e_d)}$, every input strand in $Z$ will have at least {$\lfloor \frac{K}{2} \rfloor+1$} correct copies (because $\tau<\frac{1}{2}+\frac{K \bmod{2}}{2K}$). 
However, all the other subsets (where a subset refers to all the copies that agree on the index and the data parts) would be of size at most {$\lfloor \frac{K}{2} \rfloor$}, as different strands of $Z\in \overline{\cX}_{M,L,\nodollarindlen}^{(2e_i,2e_d)}$ cannot output the same erroneous copy. Hence, the naive algorithm that selects all the $M$ subsets of size at least {$\lfloor \frac{K}{2} \rfloor+1$} will retrieve the original input successfully.
\end{IEEEproof}
 
\subsection{Constructions for  $e_d>0$}
In this subsection, we construct {$(\tau,e_i,e_d)_K$}-\NameCodes s. We start with the case of $\tau=1$. Recall that by Theorem~\ref{ThoremEd>0tau=1}, we need to construct a code $\cC\subseteq {\cX}_{M,L,\nodollarindlen}$ such that $\cD(\cC)>(2e_i,2e_d)$. In order to do so, we use a code on the data-field and then apply the constructions for the case of $e_d=0$. Let $A(n,d)$ denote the largest size of a length-$n$ binary code with Hamming distance $d$.  

\begin{lemma}\label{Lem:tau=1,e_d>0Construc}
It holds that $$A_{M,L,\nodollarindlen}(1,e_i,e_d,K)\geq {A(L-\nodollarindlen,2e_d+1)\choose M}F(\nodollarindlen,M,2e_i+1).$$   
\end{lemma}
\begin{IEEEproof}
Let
$\cC'\subseteq \{0,1\}^{L-\nodollarindlen}$ be a largest size code with Hamming distance $2e_d+1$ of size $A(L-\nodollarindlen,2e_d+1)$. For every set of $M$ different codewords from $\cC'$ (the number of such sets is $ {A(L-\log(M),2e_d+1)\choose M}$), we can apply any of the constructions for the case of $e_d=0$. This is a valid construction since for every $Z_1\neq Z_2\in \cC$, if $Z_1$ and $Z_2$ have a different data-field set, then for every permutation that matches between the strands of $Z_1$ and the strands of $Z_2$, there will be a data-field that is matched to a different data-field, and since we took all the data-fields from $\cC'$, we will have that $\cD(Z_1,Z_2)>(0,2e_d+1)$. On the other end, if $Z_1$ and $Z_2$ have the same data-field set, then we will have $\cD(Z_1,Z_2)>(2e_i+1,0)$.       
\end{IEEEproof}

Next we study the case of $\tau<1$ such that $\frac{MK}{2M-1}\leq \lfloor\tau K\rfloor$.
\begin{lemma}\label{lem:16}
For $\tau$ such that $\frac{MK}{2M-1}\leq \lfloor\tau K\rfloor< K$, we have that 
$$A_{M,L,\nodollarindlen}(\tau,e_i,e_d,K)\geq {A(L-\nodollarindlen,2e_d+1)\choose M}F(\nodollarindlen,M,e_i+1).$$

\end{lemma}
\begin{IEEEproof}
Let
$\cC'\subseteq \{0,1\}^{L-\nodollarindlen}$ be a largest size code with Hamming distance $2e_d+1$ of size $A(L-\nodollarindlen,2e_d+1)$. For every set of $M$ different codewords from $\cC'$, we can apply any of the constructions for the case of $e_d=0$, let $\cC$ denote the code obtained. First note that since all the data-fields of any codeword in $\cC$ are different and taken from a code with Hamming distance $2e_d+1$, we have that that  $\cC\subseteq\overline{\cX}_{M,L,\nodollarindlen}^{(2e_i,2e_d)}$. Thus, by Lemma~\ref{Lem:Tau>0.5ConstruFirstCond} we need to show that $\cD(\cC)>(e_i,e_d)$. This is indeed the case, and it follows in the same way as in the proof of Lemma~\ref{Lem:tau=1,e_d>0Construc}.    
\end{IEEEproof}

For $\frac{K}{2}\leq \lfloor\tau K\rfloor<\frac{MK}{2M-1}$, by Lemma~\ref{Lem:E_d>0Tau>0.5BetterCon}, we need the code $\cC$ which is obtained in Lemma~\ref{lem:16} to be in $\overline{\cX}_{M,L,\nodollarindlen}^{(e_i,e_d)}$ and not necessarily in $\overline{\cX}_{M,L,\nodollarindlen}^{(2e_i,2e_d)}$, thus we can have a better bound for this case.

\begin{lemma}
For $\frac{K}{2}\leq \lfloor\tau K\rfloor<\frac{MK}{2M-1}$, we have that 
$$A_{M,L,\nodollarindlen}(\tau,e_i,e_d,K)\geq {A(L-\nodollarindlen,e_d+1)\choose M}F(\nodollarindlen,M,e_i+1).$$

\end{lemma}

Next we study the case of $\tau<\frac{1}{2}+\frac{K \bmod{2}}{2K}$, and $\nodollarindlen=\log(M)$. In~\cite{SYLWZ20}, the authors proved that if $L$ satisfies the following equation
\begin{align}
&L-2\log(M)\geq 5(2e_d+1)+2e_i\cdot\log(\log(M))+\log\big(B_{2e_i}(\log(M))\cdot B_{2e_d}(L-\log(M))\big)+3,\label{CondtionForL} 
\end{align}
where $B_r(n)=\sum_{i=0}^r {n\choose i}$, then there exists a construction that requires a single redundancy bit and returns a code $\cC\subseteq\overline{\cX}_{M,L,\log(M)}^{(2e_i,2e_d)}$. 
 Using this and Lemma~\ref{Lem:XbarForE_d>0AndT<half} we obtain Corollary~\ref{Cor:E_d>0AndTau<0.5}. 

\begin{corollary}\label{Cor:E_d>0AndTau<0.5}
For $\tau<\frac{1}{2}+\frac{K \bmod{2}}{2K}$ and $L$ that satisfies equation~(\ref{CondtionForL}) we have that \begin{align}
A_{M,L,\log(M)}(\tau,e_i,e_d,K)\geq 2^{M(L-\log(M))-1}.\nonumber   
\end{align}  
\end{corollary}

As in the case of $e_d=0$, from a code $\cC\subseteq \cX_{M,L,\log(M)}$ which is a $(\tau,e_i,e_d)_K$-\NameCodes,  we can construct $(\tau,e_i,e_d)_K$-\NameCodes s\ for larger values of $\nodollarindlen$ by the same way as appears in Lemma~\ref{ConstructionForLargel}.   
\section{Conclusion and Future Work}
\label{sec:conclusions}
We introduced a new solution to DNA storage that integrates all three steps of retrieval, namely clustering, reconstruction, and error correction. DNA-correcting codes were presented as a unique solution to the problem of ensuring that the output of the storage system is unique for any valid set of input strands. To this end, we introduced a novel distance metric to capture the unique behavior of the DNA storage system and provide necessary and sufficient conditions for codes to be DNA-correcting codes. In our analyses we considered a variety of parameters and provided several bounds and constructions of DNA-correcting codes for the different parameter regimes.
Nevertheless, there are still several intriguing open problems that should be considered in future research. Below we present the most important ones.

\begin{enumerate}
\item For values of  $\nodollarindlen$ which are of the form  $\nodollarindlen=\log(M)+k\lceil\frac{d}{2}\rceil$ for an integer $k$, we can construct an $(\nodollarindlen,M,d)$ index-correcting codes by taking a $(\log(M),M,d)$ index-correcting code and apply Lemma~\ref{ConstructionForLargel} recursively. On the contrary, the problem of constructing $(\nodollarindlen,M,d)$ index-correcting codes when $\nodollarindlen$ does not carry this form, is still open. In particular, it will be interesting to study how to construct $(\nodollarindlen,M,d)$ index-correcting codes for these values smartly and efficiently.

\item The work presents several upper bounds on the size of a maximal \indexCode ($F(\nodollarindlen,M,d)$). An important question is for which parameters these bounds are tight. We note that example~\ref{Ex:matrix} shows that the upper bound of Lemma~\ref{lem:Singelton} is tight for $M=4,d=2$. Moreover, the characterization of all the cases in which Lemma~\ref{lem:Singelton} is obtained with equality is interesting by its own.

\item For $e_d>0$, we presented several constructions that utilize an error-correcting code on the data-field. A natural question to ask is whether there are better constructions that obtain the sufficient conditions for a code to be a $(\tau,e_i,e_d)_K$-\NameCodes?

\item Throughout this work we assumed that the number of copies for each strand is exactly the same. A more realistic model is where the number of copies can be different for different strands, for example, the case where the number of copies for each strand is distributed by the some knowen distribution. In particular, the normal and the skew-normal distributions are of high interest based on previously published works~\cite{HMG19,SOSAYY21}. An important direction is to adjust the necessary and sufficient conditions for these cases in order to guarantee a successful retrieval of the information with probability $1-\epsilon$.
\end{enumerate}

\section{Acknowledgments}
We would like to thank Prof. Tuvi Etzion for helping us analyze the DNA-Distance and Prof. Moshe Schwartz for helping us study the $\ell_\infty$ distance over the Hamming distance of the indices.

\newpage
\bibliographystyle{IEEEtran}

\end{document}